# Grain Deformation of Fractured Sandstone and Stokes Local-rotation of Material Line

Chundi Feng, Rendong Huang, Yaguang Qin*, Dongjie Yang

*School of Resources and Safety Engineering, Central South University, Changsha 410083, China*

Corresponding authors: Yaguang Qin (csuqyg@163.com)

ABSTRACT: The movement and deformation of mineral grains in rocks control the failure behavior of rocks. However, at high resolution, the physical and mechanical behavior of three-dimensional microstructures in rocks under uniaxial compression has not been characterized. Here, in suit XCT (4.6 um) has been applied to investigate the behavior of mineral grains of sandstone— movement, rotation deformation and the principle strains obtained by deformation gradient tensor constructed with three principle axial vector representation of grain, indicating that the behavior of grains between the fracture and the non-fracture zone are different. For further investigate the behavior of grain cluster, the material lines are used to obtain the Stokes local rotation, namely shear strain. The finding is that: 1) the shear strain is periodic in the radial direction. 2) on average sense, the positive shear strain and negative shear strain have local concentration features.

**1. Introduction;**

The physical and mechanical behavior of rock is determined by the deformation behavior[1,2], contact behavior[3-5] and interaction relationship of rock components[6,7] (pores, grains, cement). For example, the contact[8] and friction[9] phenomena between mineral grains are related to the formation, evolution and propagation mechanism of earthquakes[10]. At the same time, the geometric features of mineral grains affect the crack propagation[11]. For example, standard principle of local symmetry[12,13] and strain-energy density criterion[14] are applied to explain repulsive cracks[15], but geometric conditions remain lacking. Liang et al.[16] studied the deformation and crack propagation path of mineral grains and found that there is a geometric correlation between crack path deflection and mineral grain motion. In addition, the deformation, movement behavior and stress evolution of mineral grains directly reflect the formation and evolution of local deformation zones and the propagation and expansion of cracks[17-21]. The above research shows that the properties of mineral grains are correlated with the stability of rock mass engineering.

However, due to technical limitation, the research on the physical and mechanical properties of mineral grains mainly focuses on numerical simulation and two-dimensional laboratory experiments on the surface of materials. Liu, Ting and Kock[22] use DEM to simulate the effects of mineral grains of different shapes and sizes under uniaxial compression on rock deformation, strength (crack initiation strength, damage strength and peak strength)[23], and friction properties between mineral grains. Liang[16] used scanning electron microscope (SEM) to study the rock crack propagation path and mineral grain migration under the condition of medium and low strain rate loading, as well as the fine granulation and flow property of mineral grains. Sundaram[24] and his colleagues studied the deformation characteristics of granular matter around pre-cracked solids under dynamic loading by Digital Image Correlation technology (DIC), and obtained dynamic propagation patterns of different pre-cracks. Tan[25] obtained the interface separation characteristics of two-dimensional plane grains indirectly by digital speckle technique and micro-mechanics theory. Based on this, a cohesive law at the microscopic scale is established, and the cohesive strength and softening modulus of interfaces is analyzed. Maruyama[1,26] used the maker line to study the rotation and deformation properties of mineral grains on the rock surface, and found that the GBS-induced grain rotation and CPO behavior affect the seismic properties of the Earth's mantle, and further revealed the stratum response characteristics and creep mechanism of the upper mantle during earthquakes. Studies have found that grain-scale microstructure controls macroscopic fracture patterns, and some fracture phenomena of rock have a strong correlation with the plastic behavior of the material microstructure, which can affect the macroscopic strength of the rock[20,27]. At the same time, Dong[28] found that the superplastic deformation behavior of feldspar and quartz reflects the deformation mechanism of the mantle on Earth.

The above studies show that the deformation and kinematics characteristics of mineral grains have a strong correlation with rock fracture and rock mass strength. However, the above research lacks a complete three-dimensional characterization of the physical and mechanical properties of mineral grains, and cannot overcome the errors caused by the out of plane displacement of mineral grains. Therefore, using the X-ray computed tomography technique, we study and characterize the deformation, movement, rotation and stress state of mineral grains after unloading, and discuss the difference between the physical and mechanical behavior of mineral grains in rock fracture zone and non-fracture zone. At the same time, material line is constructed based on rational mechanics theory, and shear strain of CT slices is calculated by measuring local rotation angle. The relationship between rock crack propagation and grain movement were studied.

**2. Experiment**

The material considered in this research is poorly cemented sandstone, without removing any water for maintaining the microstructure of the sample, from Yunnan province in China. The diameter of sandstone grains is about 0.1mm, as shown in figure 1. Before the test, the sandstone with a 3.8 mm diameter by 7.4 mm long cylindrical sample manufactured by Focused Ion Beam (FIB), which eliminate the influence of end effect, extracted from a 50 mm diameter and 100mm long core. After the test, the mineral powder was analyzed by XRD diffraction, containing quartz, muscovite, calcite, albite and hematite, as shown in figure 2. The density, elastic modulus, and poisson's ratio of mineral components are shown in table 1.

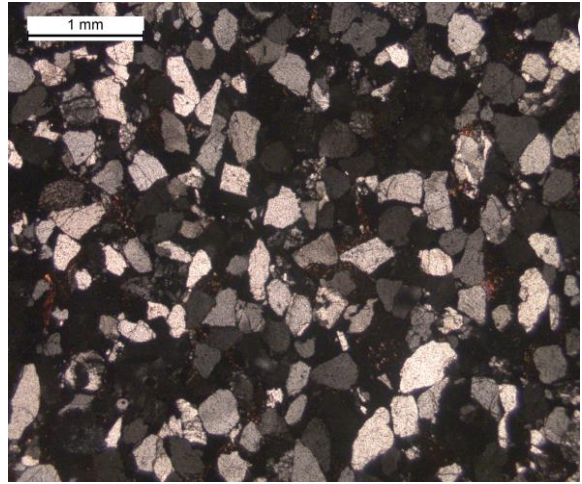

**Fig.1.** Casting image of sandstone

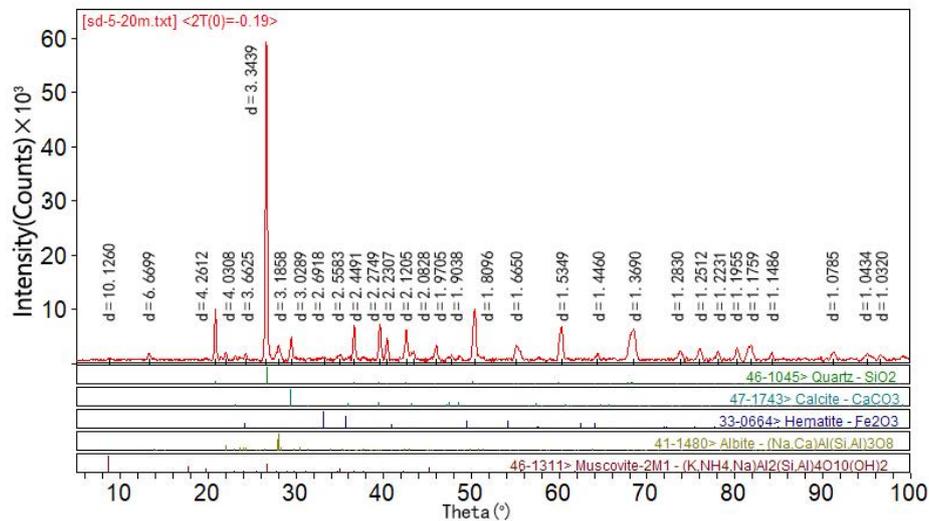

**Fig.2.** XRD patterns of sandstone sample show characteristic peaks of different minerals

**Table 1** Moduli and Density of Common Minerals in Sandstone

| Minerals | Elastic Moduli (GPa) | Poisson's Ratio | Density(g/cc) |
| --- | --- | --- | --- |
| quartz | 94.5 | 0.08 | 2.65 |
| muscovite | 56.8 | 0.25 | 2.81 |
| calcite | 84.1 | 0.32 | 2.75 |
| albite | 88.1 | 0.26 | 2.61 |
| hematite | 211.7 | 0.15 | 4.4 |

The experiment was conducted by the Xradia 520 Versa (XCT) and Material Testing Stage (MTS) with servo control system. The accuracy of the displacement and the load measurement are ±0.01mm and ±8N respectively. The maximum spatial resolution of XCT is up to 700 nm. The test process is as follows:

1) The sandstone in natural state was cut into a cylinder by FIB, and the upper and lower end surfaces of the sample were polished to eliminate friction. preservative film is used to wrap the samples to maintain the integrity of the samples after unloading.

2) The experimental devices and systems of XCT and MTS were adjusted and the sandstone was scanned for the first time. The sample was loaded with a displacement control mode with a rate of 0.003mm/min, and the stress-strain curve was recorded.

3) After unloading, the sample was scanned for the second time, during which sandstone were shot every 0.2°, uniformly rotated from 180° to -180°, and 1684 slices were taken from different angular views.

4) 3D reconstruction of the projected data before and after loading is carried out to obtain the 3D structure of its internal components, and Non-local Means algorithm is adopted to filter the images[29].

5) The de-noised image is automatically segmented by watershed algorithm[30] using software Amira, and the pixels of mineral grains are extracted for 3D reconstruction.

6) The deformation, stress and strain of mineral grains were calculated by using deformation gradient.

Figure 3 (a) and (b) show the images before and after uniaxial compression. The axial strain is 0.013, the compressive strength of rock under the uniaxial compression is 30 MPa, and the elastic modulus is 0.23GPa.

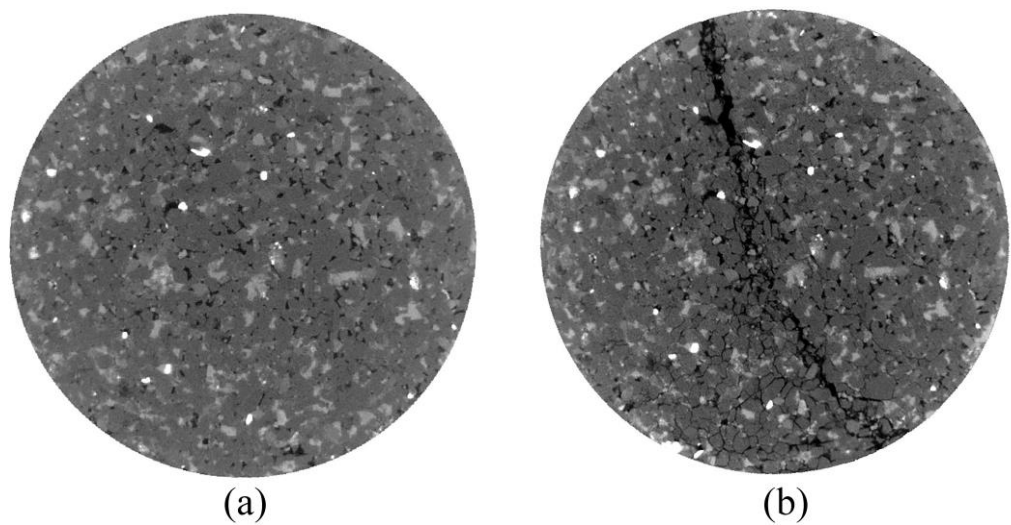

**Fig.3.** X-ray computed tomography images of sandstone: (a) before loading; (b) unloading

**3. Results and analysis**

*3.1 Deformation analysis*

To analyze movements, rotations, strain and stress of mineral grains, the radial distribution function[31] and equivalent diameter[32] of hematite grain were calculated. In order to ensure the accuracy of the calculation results, the contact ratio of mineral grains before and after compression were analyzed by calculating the radial distribution function of CT images and estimating the movements of the grains after compression. The Bounding Box algorithm were applied to extract the principal axes and volume of each pair of mineral grain, and to characterize the volume changes. Then the deformation behavior of mineral grains was analyzed by calculating its volume changes and principal axes.

*3.1.1 Deformation calculation of mineral grains based on Radial distribution function and Bounding Box algorithm.*

Radial Distribution Function (RDF) is usually used to describe grains correlation in granular system, and to acquire the order of mineral components to characterize disorder degree of three-dimensional structure of mineral grains after compression. For specified three-dimensional coordinate of mineral grains, RDF can calculate its coordinate after compression, and compute its probability distribution. Radial Distribution Function is computed by

$$g(r) = \frac{1}{\rho_0} \frac{n(r)}{V} \approx \frac{1}{\rho_0} \frac{n(r)}{4\pi r^2 \delta r} \qquad (1)$$

Where n(r) is radius of mineral grains from $r$ to $r + \delta r$, $\rho_0$ is density of mineral grains.

Equivalent radius was applied to calculate RDF, because the grains have complex shape and its irregularity make it difficult to compute RDF. Equivalent radius is computed by

$$EqD = (6 \times V_{volume} / \pi)^{1/3} \qquad (2)$$

Where *EqD* is effective radius of mineral grains, $V_{volume}$ is volume of the grains.

We then computed movements of mineral grains by calculating the changes of centre of mass of the grains before and after compression, and assign ID number to each grain. Finally, the bounding box algorithm was adopted to estimate the volume changes of all minerals before and after compression, and the boxes surrounded by them are $L_1$, $L_2$ and $L_3$ respectively.

3.1.2 Analysis and discussion on deformation results of mineral grains

Figure 4 shows the probability distribution of incremental volume of mineral grains. In general, we see that volume of mineral grains increased, but some grains decreased. According to equation (2) and the maximum volume of the grains ($14 \times 10^{-27}$mm$^3$), the maximum diameter of the grains is 0.3 mm. This grain then were placed in sample centre and used to calculate the probability distribution of the contact ratio of mineral grains in the period of *nEqD* (*N*= 1，2，3). Figure 5 shows that the *g(r)* of this grain equals 1 at the position of *EqD=1R*. The result of the *g(r)* of the rest of grains are less than 0.0769, showing that there is no contact among mineral grains. Therefore, grain tracking results before and after compression are reliable, and the it can be used for the following analysis.

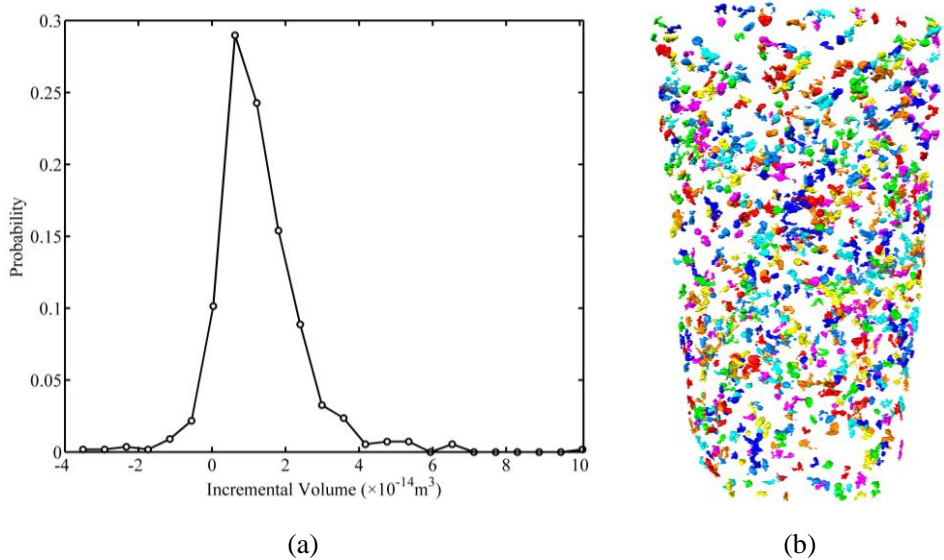

(a)            (b)

**Fig.4.** (a) Probability distribution of incremental volume of grains after unloading (b) the distribution of mineral grains after unloading

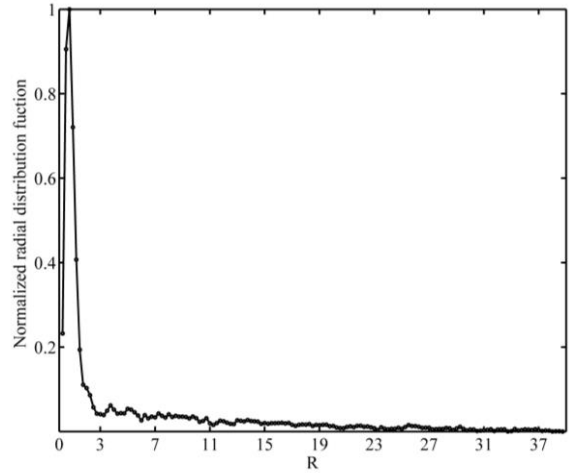

**Fig.5.** Normalized radial distribution function

Figure 6 (d) shows the concept of bounding box, and $L_1$, $L_2$ and $L_3$ represent the minimum, intermediate and maximum axes lengths of the box. The compactness was considered in the bounding box algorithm to ensure the authenticity and reliability of the results.

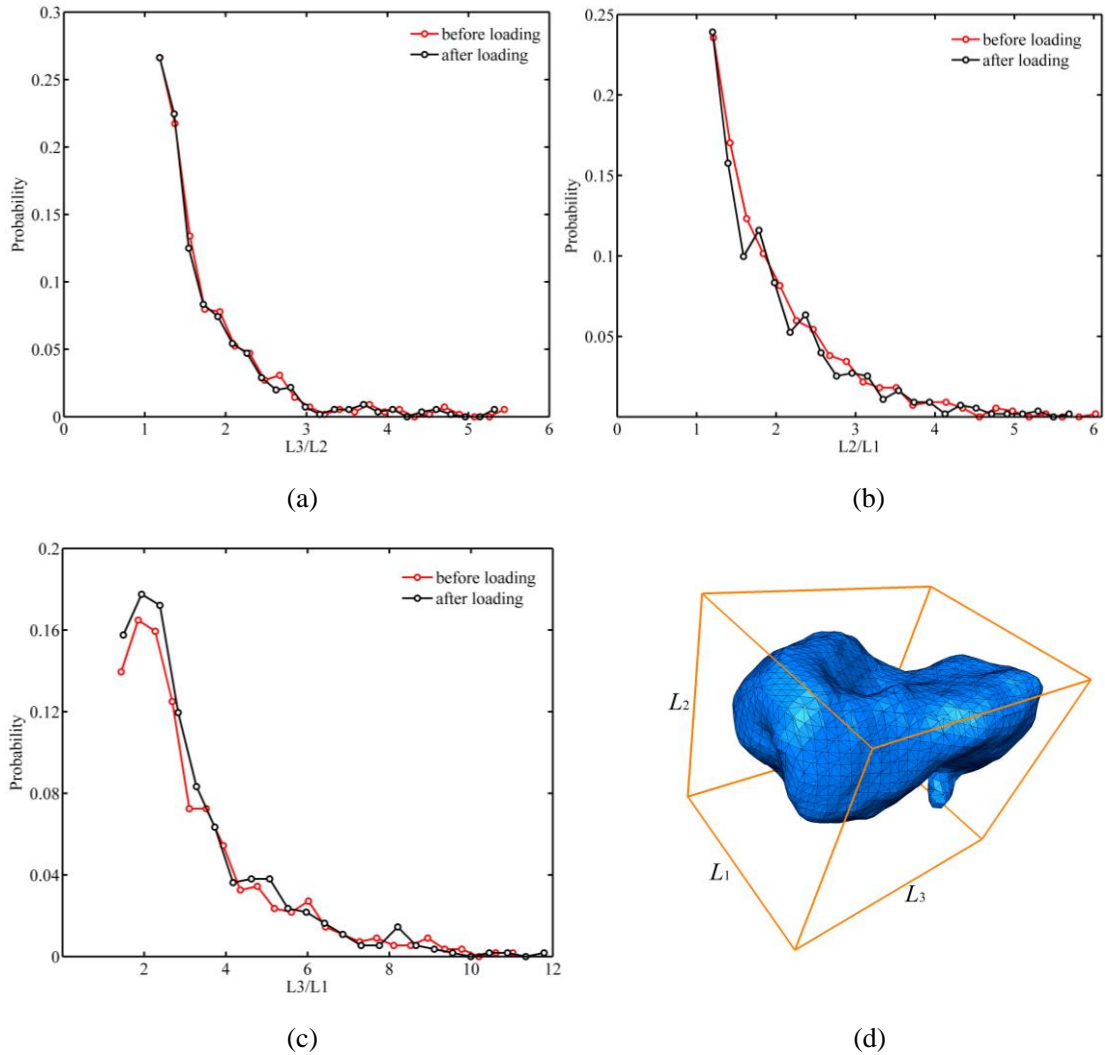

**Fig.6.** (a) - (c) are probability distribution of $L_3/L_1$, $L_3/L_1$, $L_3/L_1$ before and after loading, where $L_1$, $L_2$ and $L_3$ are the bounding box lengths respectively. (d) The bounding box.

In figure 6 (b), the two curves of $L_3$ and $L_2$ are similar in shape. Only at the point of 2.6 of $L_3/L_2$, there is a slight fluctuation, indicating that the lengths of $L_3$ and $L_2$ does not change on the whole, and the deformation trends of two curves are the same. The results show that after uniaxial compression, mineral grains have the same change trend (i.e. increase or decrease simultaneously) or remain unchanged about the lengths of maximum axes and the second major axes. According to figure 6 (d), after compression, the overall volume of mineral grains increases. Therefore, it can be judged that the length of mineral grains $L_3$ and $L_2$ increases simultaneously.

In figure 6 (c), the value of $L_2/L_1$ after compression jumps between 1.5 and 4.5 compared with that of before compression, and the fluctuation between 1 and 1.5 shows an increasing trend, followed by a decreasing trend, and gradually approaches the curve before compression. Therefore, after compression, the volume of mineral grains mainly showed an increasing trend, and the length of the $L_3$ axes of mineral grains changed greatly, which was consistent with figure 6 (b), and the fluctuation gradually decreased, which was consistent with figure 6 (c). The experimental results in figure 6 (a) - (c) are mutually verified. In addition, the volume change of mineral grains after compression may change the sphericity of the grains, so the change of grain's shape will lead to the change of contact area and contact relationship among grains, and leading to the change of friction properties of grains.

Furthermore, continue to analyze the sphericity of mineral grains, and the sphericity computed by $(36\pi V^2/S^3)^{1/3}$. Where $V$ is the volume of mineral grains, $S$ is the area of the grains. The sphericity of mineral grains before and after compression is 0.27- 0.993 and 0.333- 0.994 respectively. The mean and variance of the sphericity before and after compression are $\mu$= 0.730，$\sigma$= 0.152 and $\mu$= 0.766，$\sigma$= 0.135 respectively. More than 95% of the volume increment of the grains is mainly concentrated in the range of $0-3\times10^{-14}m^3$, as shown in figure 4. Therefore, from the statistical perspective, as the length of $L_3$ and $L_2$ axes of some mineral grains increases, the length of $L_1$ axes increases, resulting in the sphericity of most mineral grains increases.

*3.2 Translation and rotation analysis*

*3.2.1 Computational method of translation and rotation of mineral grains*

In order to extract the displacement of each pair of mineral grains, the grains before and after the compression are placed in the same coordinate system. Firstly, the first order moments were applied to calculate the centroid of the grains, so as to obtain the spatial position of them. Secondly, the centroid displacement before and after compression was calculated by using the distance formula between two points in geometry. Then, the probability distribution of the total displacement of grains was calculated, and the second order moments was applied to get the covariance matrix, $A$, computed by equation (3). For this symmetric matrix, the maximum, second and minimum eigenvalues of mineral grains were computed by normal vector decomposition[33]. Finally, the translation and rotation of 689 grains were analyzed. In this paper, the density change of mineral grains in the process of movement was ignored, and it is assumed that the grain density is uniform before and after the movement, so the centroid of mineral grains is equivalent to the center of mass.

$$A = \begin{bmatrix} \frac{1}{A(X)}\int_X xdxdydz & \frac{1}{A(X)}\int_X (x-M_{1x})(y-M_{1y})dxdydz & \frac{1}{A(X)}\int_X (x-M_{1x})(z-M_{1z})dxdydz \\ \frac{1}{A(X)}\int_X (x-M_{1x})(y-M_{1y})dxdydz & \frac{1}{A(X)}\int_Y xdxdydz & \frac{1}{A(X)}\int_X (y-M_{1y})(z-M_{1z})dxdydz \\ \frac{1}{A(X)}\int_X (x-M_{1x})(z-M_{1z})dxdydz & \frac{1}{A(X)}\int_X (y-M_{1y})(z-M_{1z})dxdydz & \frac{1}{A(X)}\int_Z xdxdydz \end{bmatrix} \quad (3)$$

*3.2.2 Results analysis for translation and rotation of mineral grains*

In figure 7 (a) shows the displacement distribution of each pair of mineral grains before and after compression, where the straight lines represent the displacement of each pair of grains. We can see the position of mineral grains before and after compression, and the movement of mineral grains mainly presents the overall sinking trend, while the opposite movement phenomenon appears in the local area. Figure 7 (a) shows the probability distribution of displacement of mineral grains after compression. The displacement of mineral grains was mainly distributed between 0.035mm and 0.2mm, among which the displacement of 22% mineral grains was concentrated at 0.06mm and the displacement of 13% mineral grains was concentrated between 0.13 and 0.14mm.

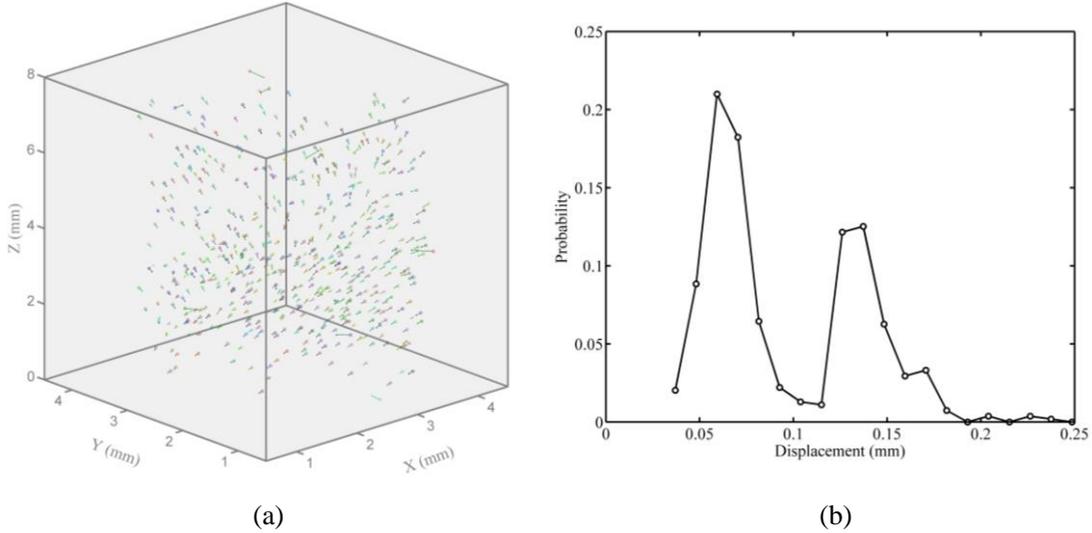

(a)          (b)

**Fig.7.** (a) The displacement for each pair of grains (b) Probability distribution of total displacement for grains.

According to the results of radial function g(r), the spatial distribution of mineral grains is discrete, and the displacement of each mineral grain is less than the distance between any two mineral grains, so as to ensure reliable registration results before and after compression, as shown in figure 7 (a). The number of grains at different normalized heights, as shown in table 2, was 227 in the range of 0.6-0.8. The minimum number of grains were 70 in the height range of 0.8-1.0. The distribution of 689 mineral grains in figure 8 (a) on the normalized height of Z-axis was analyzed. Table 2 shows the number of mineral grains located at each normalized height in figure 7 (b). Obviously, the number of mineral grains in the top position of sandstone is smaller than that in other positions. Due to the insufficient number of samples at the top of the sample, the accuracy of the results may be low, and the error of the calculation results cannot be specifically estimated.

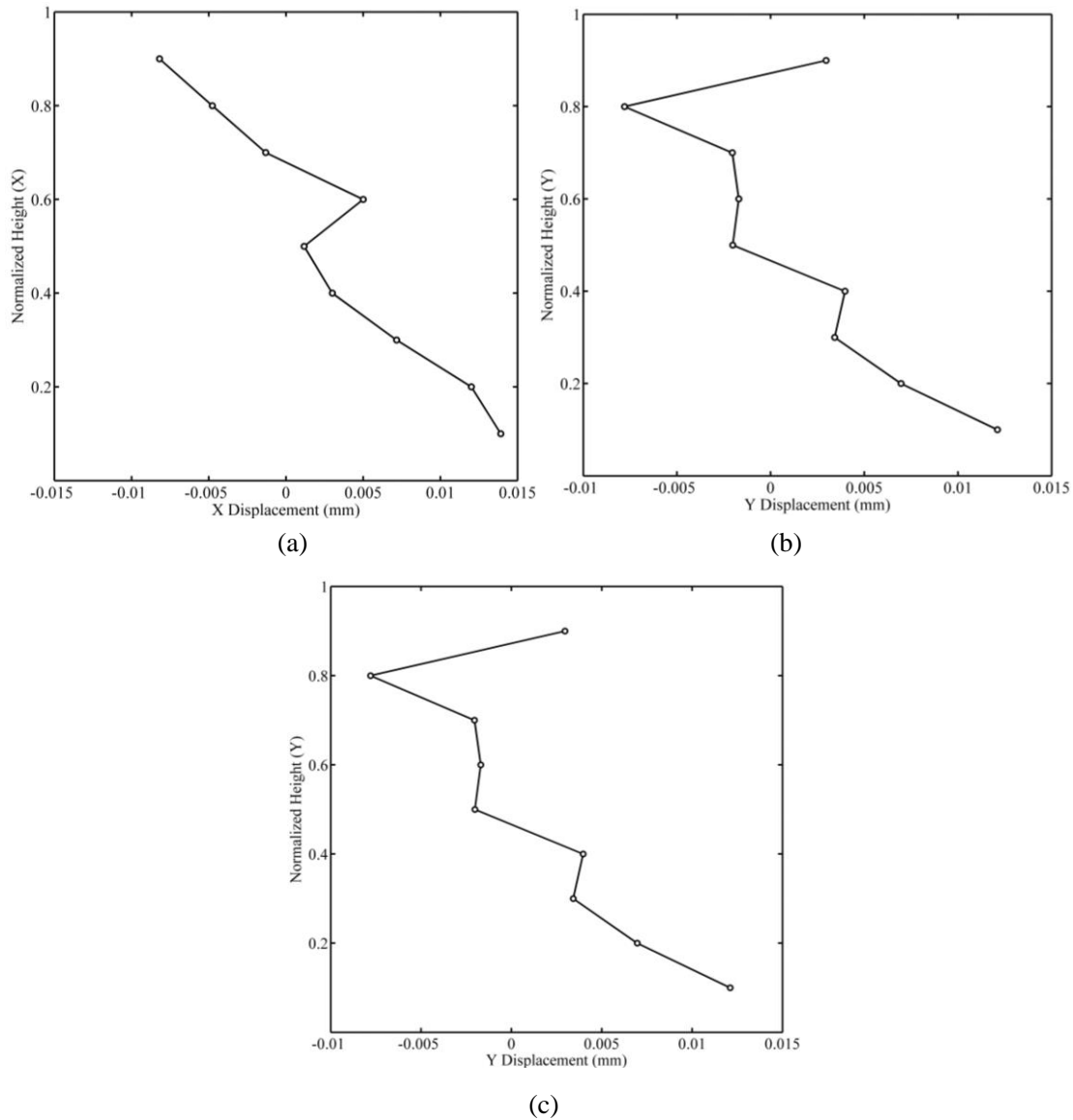

(a)          (b)

(c)

**Fig.8.** Average total displacement at each normalized height, *x*, *y* and *z* respectively.

Figure 8 (c) shows that mineral grains experienced the most total displacement at the top of the sandstone. Figure 8 (a) and (b) shows that the most displacement in horizontal plane occurred at the side of the sandstone, the displacements of X axis and Y axis are highly symmetrical. The radial displacement near the central axis is smaller, while the displacement far from the central axis is larger. This is due to the transverse expansion of the sandstone during compression, so the cumulative displacement of the grains far away from the central axis is larger.

Figure 9 (a) shows that the average rotation of mineral grains as a function of normalized height, and the rotation angle of grains is shown in figure 9 (b). The average incremental angle of the grains decreases with the increase of height, and the maximum value appears at the height of sample 0.7. The change of the average incremental angle of the upper sandstone is larger than that of the lower sandstone, which may be caused by the large friction of the lower sample. Therefore, as the height increase, the rotation angle of mineral grains tends to increase along the loading direction.

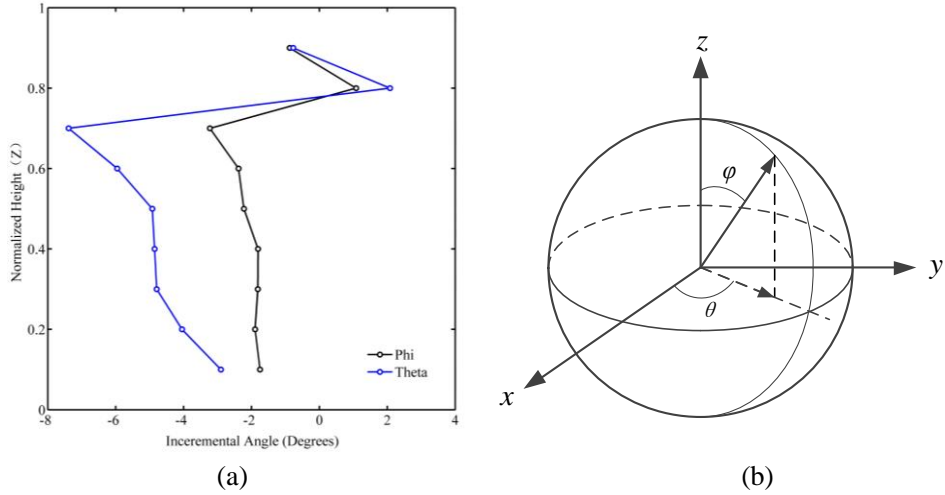

(a)                                                 (b)

**Fig.9.** The incremental angel of mineral grains as a function of normalized height (*z*).

*3.3 The principal strains and stress calculation and statistical analysis of mineral grains*

*3.3.1 Stress-strain computational method for mineral grains*

Approximating irregular grain shape by ellipsoid is a practical way to evaluating effective properties of mineral grains. We computed the covariance matrix and eigenvalues for tracked grains before and after compression as described in §3.2. By applying normal vector decomposition, this symmetric matrix can be described as $C=Q \wedge Q$, where $Q$ represents the eigenvectors' matrix and $\wedge$ represents eigenvalues' matrix, which can be used to describe three axes of each grain[33]. Then before deformation, the grain is expressed as:

$$A^{0i}_j \mathbf{e}_i = \mathbf{E}_j \tag{4}$$

After deformation, this grain is expressed as:

$$A^{i}_j \mathbf{e}_i = \mathbf{E}_j \tag{5}$$

Where $e_i$ is unite vector, and $A^{0i}_j$ and $A i_j$ represent the transform matrix of each pair of grain before and after loading, and $E_j$ is eigenvalues of the grain. Then we computed the deformation tensor, $F i_j$, by

$$A^{i}_j = F^{l}_j A^{0i}_l \tag{6}$$

Strain and stress tensor, $\varepsilon i_j$ and $\sigma i_j$, computed by

$$\varepsilon^{i}_j = F^{i}_j - \delta^{i}_j \tag{7}$$

$$\boldsymbol{\sigma}_{ij} = \lambda \varepsilon_{ll} \delta_{ij} + 2\mu \varepsilon_{ij} \tag{8}$$

Where $\delta_{ij}$ is Kronecker symbol, $\lambda = \dfrac{E\upsilon}{(1+\upsilon)(1-2\upsilon)}$ and $\mu = \dfrac{E}{2(1+\upsilon)}$ are Lame constants, E and $\upsilon$ are Young's modulus and Poisson ratio respectively.

*3.3.2 Results analysis on stress-strain distribution of mineral grains*

Figure 10 illustrates after compression the average stress and strain components of mineral grains as a function of normalized height, (h-z)/h, where h is the height of the sandstone and z is the z coordinate of grain centroid. On the whole, the curves of principal components of strain and stress of the grains are consistent in shape.

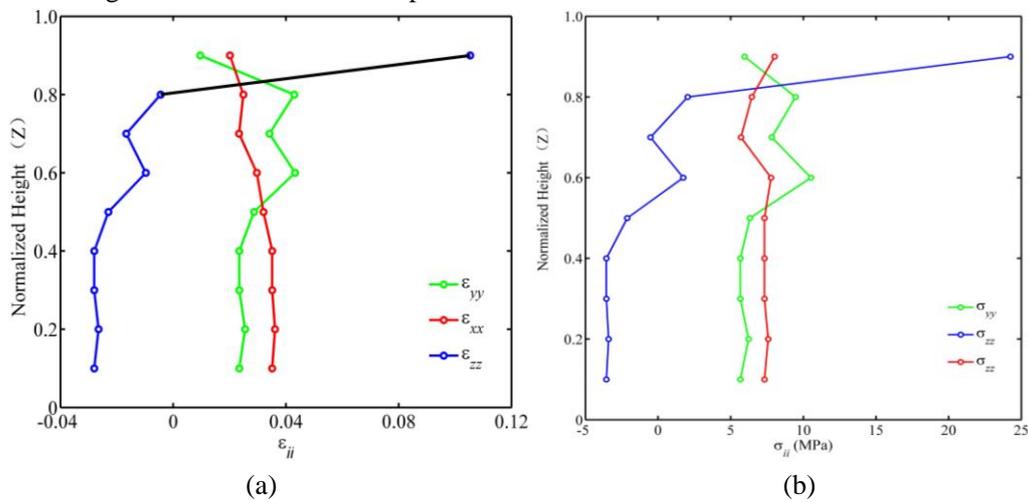

**Fig.10.** (a) and (b) are average strain and stress components of grains as a function of normalized height (z) after unloading.

In figure 10 (a), strain components of mineral grains, $\varepsilon_{xx}$, $\varepsilon_{yy}$ and $\varepsilon_{zz}$, is between the range of 0.02- 0.04, 0.01- 0.04 and -0.3- 0.11 respectively. The fluctuation of the strain component, $\varepsilon_{zz}$, is most significant, and the fluctuation of the strain component, $\varepsilon_{xx}$, is least significant. The strain components of the mineral grains, $\varepsilon_{xx}$ and $\varepsilon_{yy}$, fluctuated within a range of 0.2-0.5, showing a certain similarity in the trend and value. The results illustrate that the average deformation of mineral grains shows the same trend in the vertical direction. The mean value of strain component, $\varepsilon_{zz}$, is mainly concentrated around -0.025 and reaches the maximum value at the bottom of the sample. It can be seen that at the bottom of the sample, the grains bear the maximum compression.

Furthermore, in the range of the sample height between 0.8 and 1, the main strain components of the mineral grains fluctuated greatly in value. This is due to the concentration of mineral grains near the fracture zone. The curve showed a maximum value at the height of 0.9, which was mainly because the grains in the fault zone were fractured, as shown in figure 11. Therefore, the mineral grains in the fault zone are subjected to large plastic deformation and even fracture, resulting in a large stress concentration, and the figure 10 shows a behavior of mutations.

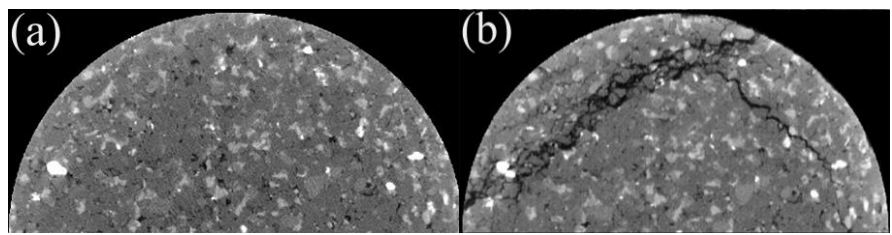

**Fig.11.** Grayscale images before and after loading at 0.9 normalized height: (a) before loading; (b) after loading

Figure 12(a) to (c) show the spatial distribution of principal strain components and principal stress components. Principal strain components, $\varepsilon_{xx}$, $\varepsilon_{yy}$ and $\varepsilon_{zz}$, range between -0.5 and 0.5, mainly concentrated between -0.1 and 0.1, as shown in figure 12(a) to (c). This phenomenon of large strain span is mainly reflected in the spatial distribution, namely fault zone and non-fault zone.

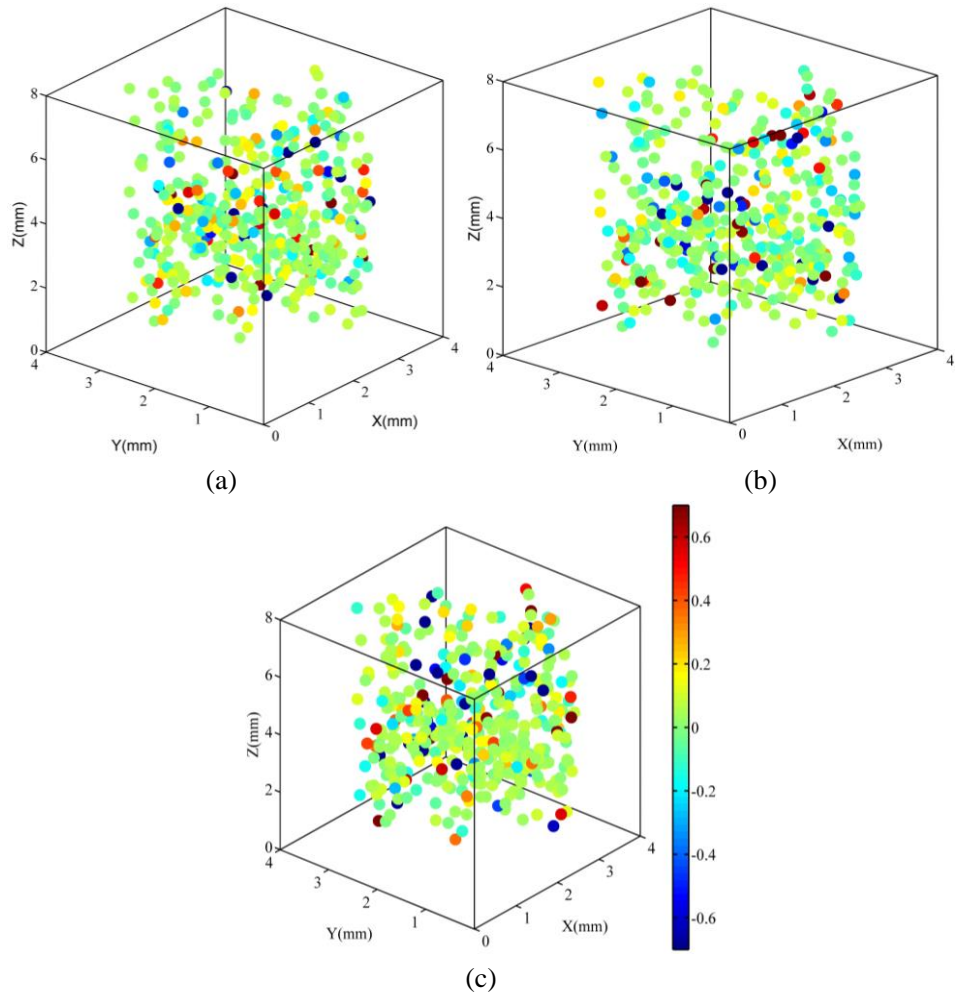

(a)　(b)

(c)

**Fig.12.** The spatial distribution of $\varepsilon_{xx}$, $\varepsilon_{yy}$, and $\varepsilon_{zz}$.

Figure 12(a) to (c) show that the area with negative values means that the compression strain constitutes an inclined plane with an angle of inclination about 45°, which is the fault surface after compression, as shown in figure 13. After compression, the strain components in the fault zone were about 5-10 times of that in other areas, and the strain in the fault zone was mainly compression strain. Meanwhile, the strain components at the top of the sample are generally larger than that at the bottom of the sample. The samples showed compression deformation as a whole, among which the compression strain in the non-fault zone was mainly concentrated between 0 and 0.1, and that in the fault zone was mainly concentrated between -0.2 and -0.4. However, the strain behavior in the local area is shown as stretch, where the tensile strain value is concentrated around 0.2. Therefore, there are complex strain behavior and stress response in sandstone.

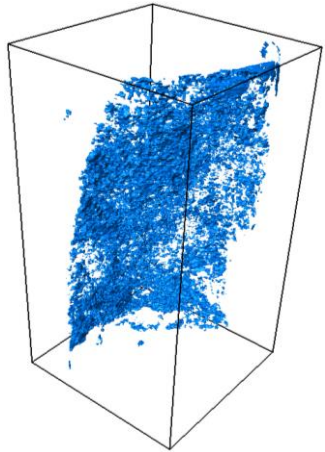

**Fig.13.** The shape of fracture

*3.3.3 Probability density distribution of spatial stress-strain of mineral grains*

Figure 14(a) and (b) show the probability distribution of strain and stress components. Figure 14(c) show the probability distribution of the sum of principal strain components, mainly ranged between -1.5 and 1.5. It is estimated that 70% of the strain concentrated between -0.5 and 0.5, indicating that the sum of the principal strain components shows a wide range, and the strain distribution is relatively concentrated. However, the value of the strain is still different from the macroscopic strain of sandstone.

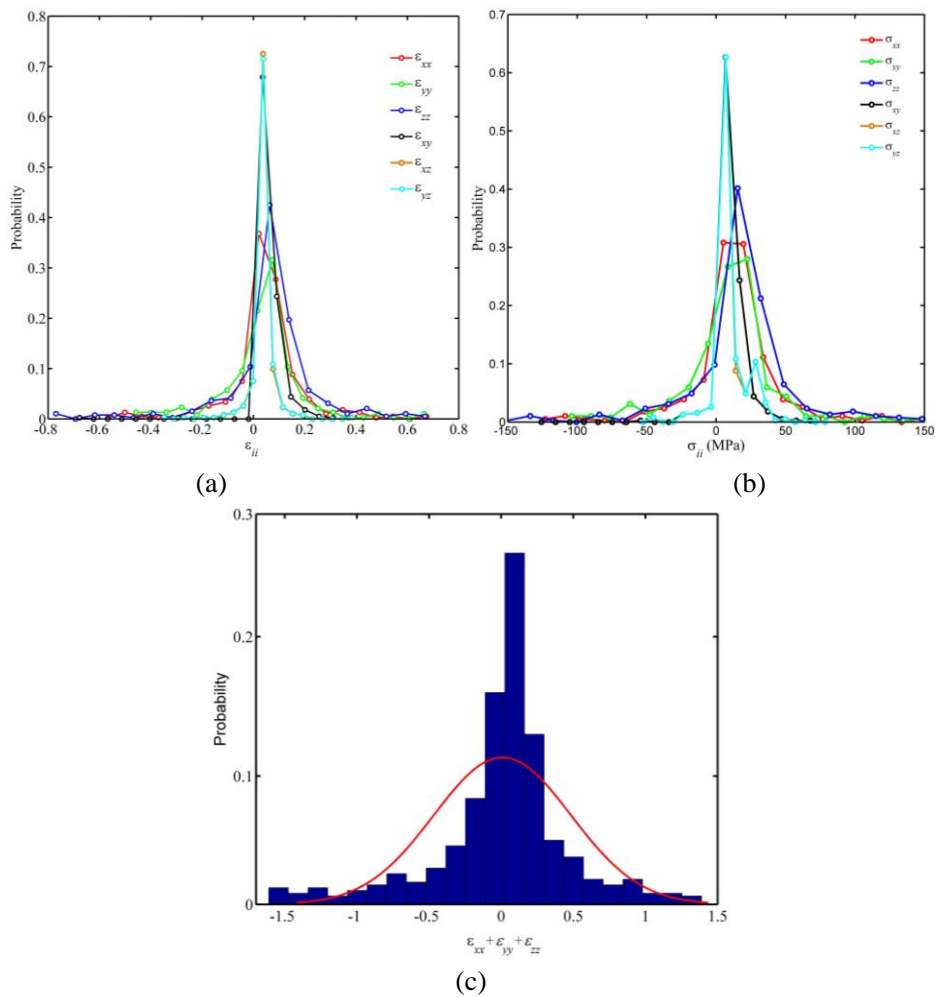

**Fig.14.** (a) and (b) are probability distribution of strain and stress components. (c) probability

distribution of the sum of strain components.

Figure 14(a) shows that the mean value of each principal strain component is 0.01, 0.13 and -0.17 respectively. Figure 14(b) shows that the mean value of stress is 3.5MPa, 23.2 MPa and -32.3 MPa respectively. The variance of each principal strain components is 0.0014, 0.0054 and 0.0023 respectively, and the variance of each stress components is 0.38, 1.20 and 0.54. The stress components of mineral grains range from -100Mpa to100MPa, mainly concentrated between -20 MPa to 30 MPa. Meanwhile, the strain components of the grains range from -0.6 to 0.6, mainly concentrated between -0.1 to 0.15. After uniaxial compression test, the uniaxial compressive strength of sandstone was calculated to be 30 MPa, the strain was 0.013, and the elastic modulus and poisson's ratio were 0.23GPa. Figure 14(b) shows the large stress on mineral grains after compression. The mean values of $\sigma_{zz}$ are negative, and the mean values of $\sigma_{xx}$ and $\sigma_{yy}$ are positive, indicating that the grains were subjected to compressive stress in the z-axis and tensile stress in the XY plane. The results are consistent with figures 12 (a) to (c). Therefore, the strain of mineral grains is much larger than the strain of sandstone, and the large plastic deformation can be found in sandstone. It is estimated that the plastic strain of the grains in the fault zone is about 30 times of the macroscopic strain of sandstone, and that of the grains in the non-fault zone is about 5 times of the macroscopic strain of sandstone, indicating that the deformation of mineral grains is much larger than the macroscopic strain of sandstone.

*3.4 Shear strain measurement*

In chapter § 3.1 to 3.3, the deformation behavior of single mineral grain after unloading was discussed. In this section, the deformation behavior of mineral grain aggregate, namely grain cluster, was studied. Theoretically, shear deformation is decomposed as two additive parts, Green shear strain and Stokes local rotation[34]. For multi-components sandstone, Green shear strain cannot be obtained as the mineral grains have varied shape, size, and stochastic distribution. Hence, the material lines[34] are used to obtain the Stokes local rotation. For further observe the complex strain of *xy* plane, *S+R* decomposition theorem[35] in Lagrange coordinate system is introduced to reveal shear strain complexity in sandstone under uniaxial compression. Based on the principle of material invariance, we used material lines to characterize the element deformation, and the shear strain contour map after fracture is acquired by tracing material lines inside the rock.

*3.4.1 Theory of material line construction on planar*

For rock samples, if a material line was constructed along characteristic components, it is generally a curve. Before deformation, the material line is expressed as:

$$ds = |ds| \cdot (\cos\alpha_0 \mathbf{e}_1 + \sin\alpha_0 \mathbf{e}_2) \tag{9}$$

where $\alpha_0$ is the angle between the starting normal line and the ending normal line. $\mathbf{e}_1$, $\mathbf{e}_2$ are the tangent vector and the normal vector of the unit length of the material line of the starting point, respectively. After deformation, this material line is expressed as:

$$ds = |ds| \cdot (\cos\alpha \mathbf{e}_1 + \sin\alpha \mathbf{e}_2) \tag{10}$$

shear strain:

$$\omega = \frac{\alpha - \alpha_0}{|ds_0|} = \theta \tag{11}$$

For studying the quartz sandstone, Equations (1)- (4) are used in this paper to practically measure tensile strain and shear strain due to the random distribution of the microstructure. Assuming that it is the planar isotropic distribution, then $S_{11}=S_{22}$. Based on the experimental data, the mean area of Green strain is assumed to be zero reasonably, namely $\varepsilon_{11} \approx \varepsilon_{22} \approx 0$. Because the independence of shear strain, they can be discarded. For plane deformation, there is a simple form of deformation tensor:

In the non-fracture zone, $\theta$ can be measured by the comparison of $ds_0$ and $ds$. $|ds_0|$ is taken as the constant value and $|ds_0|=1$ is defined as the unit scale in global measurement. For the fracture zone, since a material line is broken into two staggered material lines, the theoretical formula is adopted

$$S_1^1 = S_2^2 = 1 - \cos\theta \tag{12}$$

where $\theta$ is calculated in the fracture zone. In mechanics, the residual shear strain $\theta$ in the fracture zone is slightly smaller than the plastic shear strain $\theta$ in the non- fracture zone.

In this paper, 352 and 364 material lines in two scales were randomly selected on the CT slice. The middle point of the material line before the rock fracture is taken as the marker to draw the contour map of shear strain, as shown in figure 15. In figure 15(a) and (c), two curves were composed of five material points, and material lines were constructed using the right-hand rule. The basic principles of material line selection are as follows:

(1) The bending of the material line before deformation and after deformation is unidirectional, and the length is greater than or equal to the selected unit length, the substance on the material line must be the same.

(2) When the material line is periodic, the length of the material line should be less than half of the spatial wavelength based on the sampling law.

(3) When material lines are randomly distributed and bent, continuous data on the plane should be guaranteed in open covering of unit length.

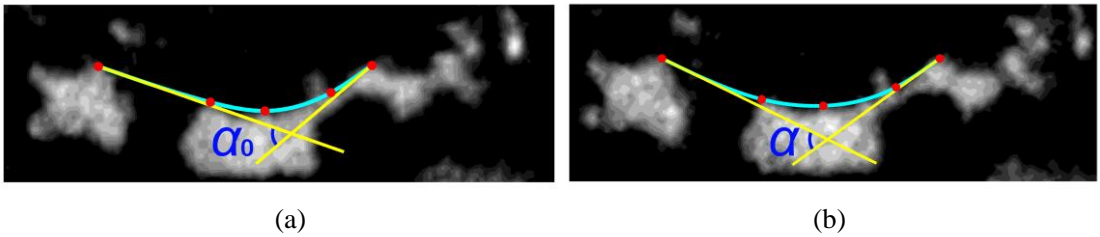

(a)　　　　　　　　　　　　　　　　(b)

Fig.15 material line: (a) before deformation; (b) after deformation.

*3.4.2 Contour maps distribution of shear strain based on material line*

Figure 16 shows contour maps of shear strain. Material lines are the proportional scale, and lines that crossed the center of the circle are the starting lines for the measuring shear strain. The shear strain was measured by two length scale lines, and the shear strain distribution was basically consistent. As shown in figure 16 (a), the maximum shear strain is 0.65 and the minimum -shear strain is -0.5. As shown in figure 16 (b), the maximum shear strain is 0.35 and the minimum -shear strain is -0.36. However, in the sub-fracture zone, the contour map of two scales showed great differences.

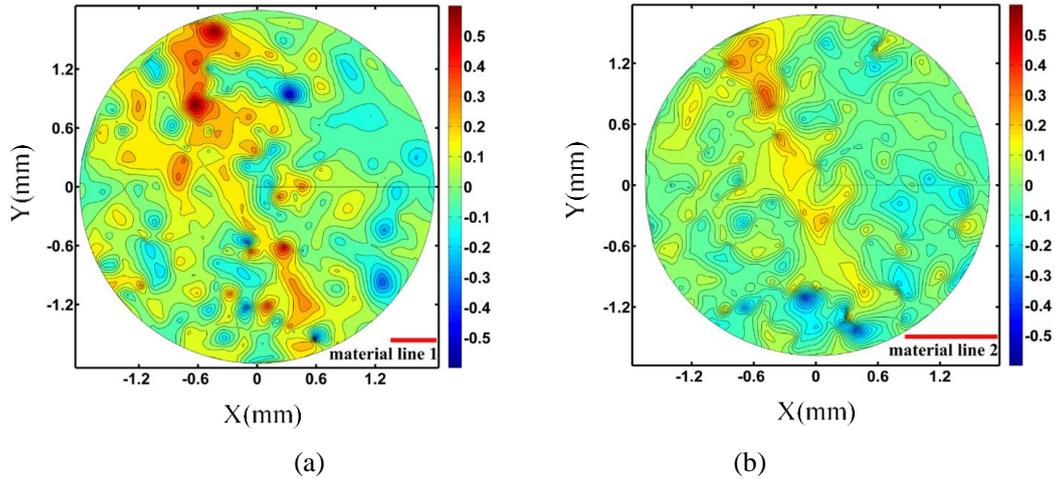

**Fig.16.** shear strain contour map: (a) measured by using material line 1; (b) measured by using material line 2.

In figure 16 (a), the critical value, $\theta$, is predicted by the contour map to be 0.5, whereas the strain of sandstone is 0.015. The shear strain in fracture zone is greater than the compression strain of sandstone, so the mechanism of fracture is that there is strain concentration near the fracture zone, namely shear concentration.

Then, near main fracture, shear strain concentration occurs in sandstone, and the strain value is proportional to the crack width. The shear strain distribution presents periodic changes, decreasing along the vertical direction of the main fracture and expending to both sides. In addition, at both ends of the main fracture, two approximately circular high shear strain zones are formed.

Furthermore, in the sub-crack network, the strain state is complex, presenting a positive and negative interphase regional distribution. The shape and size of strain islands are consistent with the grain size in CT image after unloading. However, the map measured by the large-scale material lines did not provide enough detailed information, as shown in figure 16 (b), indicating that strain measurement is related to the selection of appropriate measurement scale.

3.4.3 Shear strain profiles of contour map

To visualize numerical changes on the contour map of two scales and further explain the relationship between the periodic shear strain and its spatial position, the shear strain profiles at different angles (0°, 45°, 90°, 135°) were drawn counter-clockwise at intervals of 45 degrees by using the starting line to cut the contour map, as shown in figure 17(a)-(d) and (e)-(h). The results show that the shear strain distribution presents a periodic phenomenon of radial space. The value rises and falls periodically with the X-axis, and presents an approximate symmetry with the Y-axis. Four sets of curves are similar in shape. The profiles measured by the small-scale material lines can be divided into three regions, as shown in figure 17 (a)-(d).

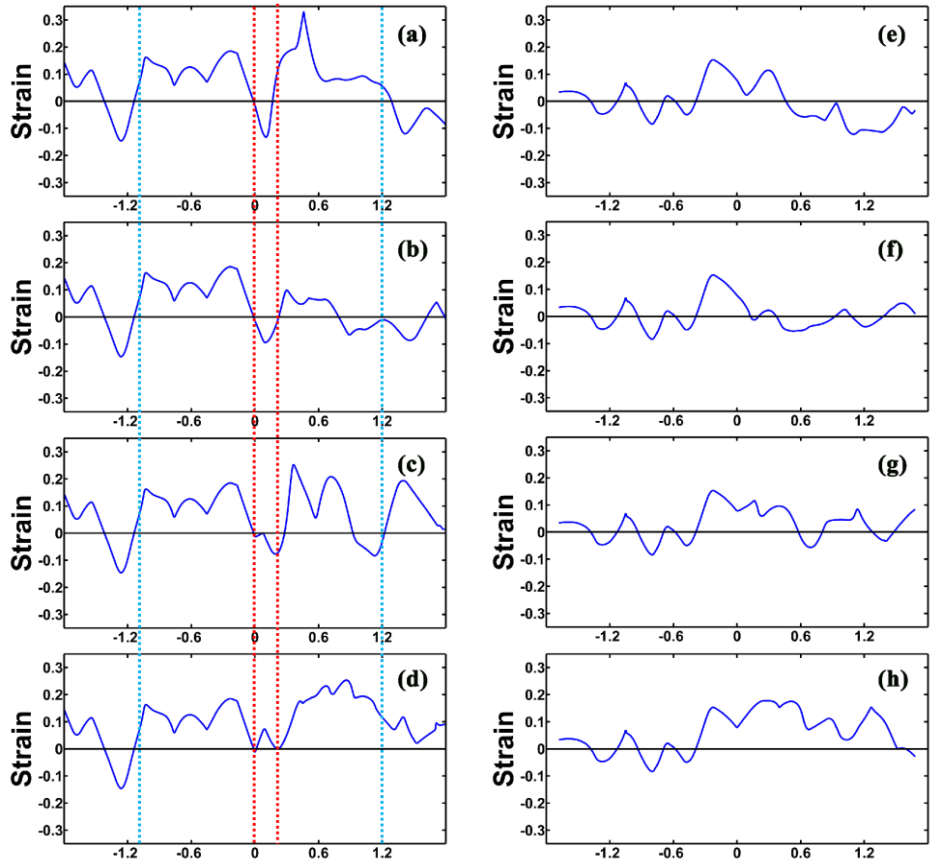

**Fig.17.** Shear strain profiles at different angles: (a) and (e) present 0° (b) and (f) present 45°, (c) and (g) present 90°, (d) and (h) present 135° with starting line.

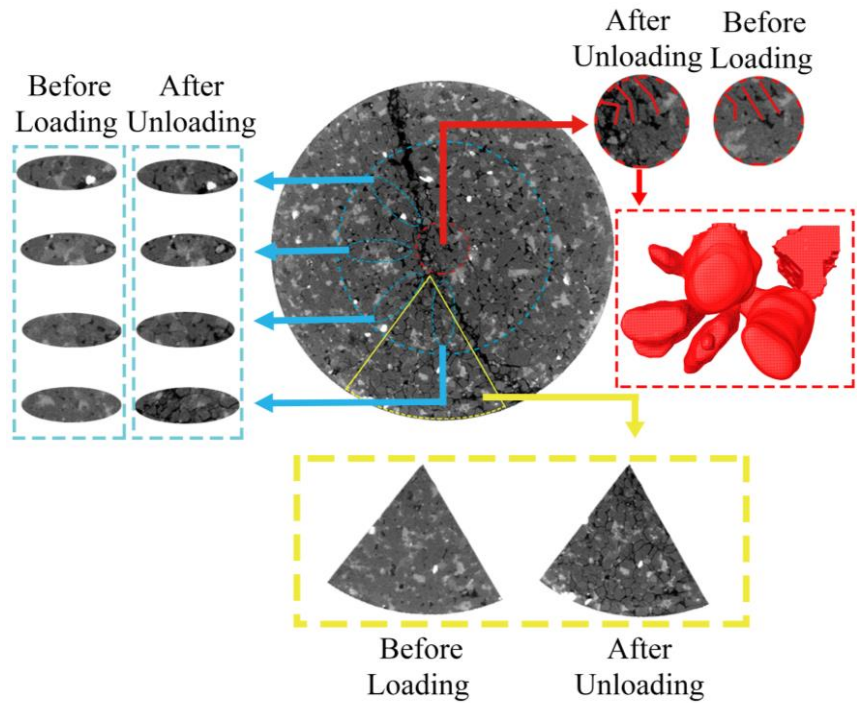

**Fig.18.** Rotations and deformation of grain cluster at the three regions

1) Center circle: this region is distributed between 0 - 0.2, as shown in the figure 17. Four sets of curves, (a)-(d), are similar in shape, range and value, corresponding to the red area in figure 18. The rotation of the central axis of mineral grains in the sub-fracture zone in this region can be explained by the grain rotation in figure 9 (a), and the axial direction of quartz grains is marked by lines. In this area, quartz grains rotate and show symmetry along the sub-crack zone, which is similar to twins. The propagation and coalescence of micro-cracks are caused by the rotation of mineral grains.

2) Middle ring: this region is distributed between -1.1 - 0 and 0.2 - 1.2, as shown in the figure 17, corresponding to the blue area in figure 18. Four sets of curves are similar in shape, but there are large differences in values and ranges. This is because the shear strain on the right side of the fault is smaller than that on the left side of the fault. The "M" shaped curve in the figure is caused by uneven rotation gradient of the cluster, as shown in figure (a) to (d), and the rotation of grain cluster has shown in figure 18(blue circles).

3) Sandstone edge: this region is distributed between ≤ -1.1 and ≥ 1.2, as shown in the figure 17, corresponding to the rest of area in figure 18. Four sets of curves have obvious differences in shape, range and value, and there is no obvious symmetry. In figure 16(b), shear strain of grain cluster in this region presents positive and negative distribution, especially in the sub-fracture zone. This is due to the contrary rotation direction of grain cluster during uniaxial compression, which results in the initiation, propagation and coalescence of cracks.

The shear strain measurement of CT slices by using two-scale material lines shows that the distribution of rock deformation field is not uniform, and complex large deformation measured by the small-scale material lines. Meanwhile, on the large scale, namely the macro scale, it can be regarded as a uniform deformation field, and a simple small deformation. The problem of large deformation and small deformation discussed here depends on the size of the sample and the amount of deformation. In this paper, the material line is used to measure the local rotation angle of the grain cluster inside the rock to overcome the error caused by the out of plane displacement of mineral grains. However, due to the problem of image accuracy and material point recognition rate, a certain error is caused, and the error cannot be estimated.

By measuring the local rotation angle of the material line, it is found that there is periodic strain distribution inside sandstone after unloading, and the residual strain is a cumulative process. After unloading, the residual shear strain still has a large value, indicating its irreversibility. Therefore, in the process of material fracture, crack initiation and propagation are generated by irreversible local rotation angle, which is represented by rotation angle parameters and the direction of rotation. When the local rotation angle parameter increases to the critical value, fracture will occur in the rotation direction. Meanwhile, for the local rotation angle, $\Theta = \Theta_c \delta(x - x_0, y - y_0, z - z_0)$, inside the medium, the point $(x_0, y_0, z_0)$ is a singularity (fracture point or defect). In theory, it satisfies the equation $\nabla^2 \Theta = 0$ (Laplace equation) in the elastic deformation zone $\Theta$, presenting a circular fatigue striation centered on the singularity.

The results show that there is a characteristic deformation scale, $L_0$, determined by the microstructure of material. When the scale $L<<L_0$, or $L>> L_0$, the shear strain is very small, and Moreover, the periodic distribution of shear deformation in space shows that the characteristic deformation scale, $L_0$, is a half-period. The shear strain scale of the fatigue fracture deformation may be the key data for improving the microstructure of materials to improve the fatigue fracture resistance. For most fatigue and fracture, the maximum value of local rotation angle is distributed in bands, corresponding to the shear failure zone in engineering.

**4. Conclusion**

Rock is a heterogeneous material with complex internal micro-components and structures. According to the difference between the mechanical properties of rock micro-components and the macro-mechanical properties of rock, this paper uses XCT and digital image processing theory to study the migration and fracture characteristics of rock microscopic components, and provides a new perspective to study the internal microstructure of rock. It is of great significance to understand the internal structural characteristics and macro physical and mechanical properties of rocks. At the same time, it is of great significance to realize the three-dimensional microscopic geometric modeling of rock, quantitatively characterize rock geometric model and rock fracture failure analysis. The main conclusions are as follows:

(1) After unloading, mineral grains in the non-fault zone move radially outward, while the volume of mineral grains decreases along its maximum principal axis and the sphericity increases. The deformation of mineral grains in the fault zone is large and their volumes increase. The strain and stress states are calculated by the deformation of the grain principal axis. The results show that the internal grains of the sandstone exhibit complex stress states and deformation behaviors.

(2) By analyzing the spatial distribution characteristics of the principal strain components of grains, the results show that there are significant differences in the deformation behavior of grains between the fracture zone and the non-fracture zone. The grain strain components $\varepsilon_{xx}$, $\varepsilon_{yy}$ are significantly smaller than $\varepsilon_{zz}$ and the fluctuation degree of $\varepsilon_{xx}$ and $\varepsilon_{yy}$ is less than $\varepsilon_{zz}$. grains are subjected to compressive stress in the Z-axis direction and tensile stress in the XY plane. The grain strain in the fracture zone and non-fracture zone is about 30 times and 5 times of the macro strain of the sample respectively, which indicates that there is a significant difference of deformation behavior between the grains and the macro sample.

(3) The shear strain profiles are plotted. The results show that there is a near-periodic shear strain distribution phenomenon in the radial direction. Through the analysis of CT gray scale image and contour profile of rock, it can be concluded that the shear strain near the fault zone is obviously larger than the compressive strain. It shows that the basic mechanism of fracture is the shear stress concentration near the fracture zone. In the fracture process of materials, the initiation and propagation of cracks are caused by irreversible local rotation angles. When the local rotation angle increases to the critical value, cracks will occur in the rotation direction.

**Acknowledgements**

This work was supported in part by the National Key Research and Development Program of China during the Thirteenth Five Year Plan Period: The Continuous Mining Theory and Technology on Spatiotemporal Synergism of Multi-Mining Areas Within a Large Ore Block for Deep Metal Deposit under Grant 2017YFC0602901, and in part by the Fundamental Research Funds for the Central Universities of Central South University under Grant 2017zzts204.

**References**

1. Maruyama G, Hiraga T. Grain- to multiple-grain-scale deformation processes during diffusion creep of forsterite + diopside aggregate: 2. Grain boundary sliding-induced grain rotation and its role in crystallographic preferred orientation in rocks. *J Geophys Res*. 2017;122(8):5916-5934.
2. Austrheim HK, Dunkel KG, Plümper O, Ildefonse B, Liu Y, Jamtveit BR. Fragmentation of wall rock garnets during deep crustal earthquakes. *Sci Adv*. 2017;3(2):e1602067.
3. Emeriault F, Cambou B. Micromechanical modelling of anisotropic non-linear elasticity of granular medium. *Int J Solids Struct*. 1996;33(18):2591-2607.
4. Shen Z, Jiang M, Thornton C. DEM simulation of bonded granular material. Part I: contact model and application to cemented sand. *Comput Geotech*. 2016;75:192-209.
5. Sufian A, Russell AR. Microstructural pore changes and energy dissipation in Gosford sandstone during pre-failure loading using X-ray CT. *Int J Rock Mech Min*. 2013;18:2591-607.
6. Putnis A, Mauthe G. The effect of pore size on cementation in porous rocks. *Geofluids*. 2001; 1(1): 37-41.
7. Li X, Lin C, Miller JD, Johnson WP. Pore-scale Observation of Microsphere Deposition at Grain-to-Grain Contacts over Assemblage-scale Porous Media Domains Using X-ray Microtomography. *Environ Sci Technol*. 2006;40(12):3762-3768.
8. Madadi M, Saadatfar M. A finite-element study of the influence of grain contacts on the elastic properties of unconsolidated sandstones. *Int J Rock Mech Min*. 2017;93:226-233.
9. Li Q, Tullis T E, Goldsby D, Carpick R W. Frictional ageing from interfacial bonding and the origins of rate and state friction. *Nature*, 2011, 480(7376):233.
10. Verberne BA, Plumper O, de Winter DA, Spiers CJ. Rock mechanics. Superplastic nanofibrous slip zones control seismogenic fault friction. *Science*. 2014;346(6215):1342-1344.
11. Bobylev SV, Ovid'Ko IA. Grain boundary rotations in solids. *Phys Rev Lett*. 2012;109(17):175501.
12. Cotterell B, Rice J R. Slightly curved or kinked cracks. *Int J Fracture*. 1980,16(2):155-169.
13. Gol'Dstein R V, Salganik R L. Brittle fracture of solids with arbitrary cracks. *Int J Fracture*. 1974;10(4): 507-523.
14. Sih G C. Some basic problems in fracture mechanics and new concepts. *Eng Fract Mech*. 1973;5(2):365-377.
15. Ghelichi R, Kamrin K. Modeling growth paths of interacting crack pairs in elastic media. *soft matter*. 2015;11(40):7995-8012.
16. Liang C, Wu S, Li X. Research on Micro-Meso Characteristics of Granite Fracture Under Uniaxial Compression at Low and Intermediate Strain Rates. *Chin J Rock Mech Eng*. 2015;34(s1):2977-2986.
17. Hurley RC, Lind J, Pagan DC, Akin MC, Herbold EB. In situ grain fracture mechanics during uniaxial compaction of granular solids. *J Mech Phys Solids*. 2018;112:273-90.
18. Sibson R H. Earthquakes and rock deformation in crustal fault zones. *Annu Rev Earth Pl Sc*. 1986;14(1):149-175.
19. Hurley R C, Hall S A, Wright J P. Multi-scale mechanics of granular solids from grain-resolved X-ray measurements. *P Roy Soc A-Math Phy*. 2017;473(2207):20170491.
20. Hurley RC, Lind J, Pagan DC, Homel MA, Akin MC, Herbold EB. Linking initial microstructure and local response during quasistatic granular compaction. *Phys Rev E*. 2017;96(1):12905.

21. Lan H, Martin CD, Hu B. Effect of heterogeneity of brittle rock on micromechanical extensile behavior during compression loading. *J Geophys Res*. 2010;115(B1):B01202.
22. Kock I, Huhn K. Influence of particle shape on the frictional strength of sediments — A numerical case study. *Sediment Geol*. 2007;196(1-4):217-233.
23. Liu G, Rong G, Peng J, Hou D, Zhou C. Mechanical behaviors of rock affected by mineral particle shapes. *Chin J Geotech Eng*. 2013;35(3):540-550.
24. Sundaram BM, Tippur HV. Full-field measurement of contact-point and crack-tip deformations in soda-lime glass. Part-II: Stress wave loading. *Int J Appl Glass Sci*. 2018;9(1):123-136.
25. Tan H, Liu C, Huang Y, Geubelle P. The cohesive law for the particle/matrix interfaces in high explosives. *J Mech Phys Solids*. 2005;53(8):1892-1917.
26. Maruyama G, Hiraga T. Grain- to multiple-grain-scale deformation processes during diffusion creep of forsterite + diopside aggregate: 1. Direct observations. *J Geophys Res*. 2017;122(8):5890-5915.
27. Ting J M, Meachum L, Rowell J D. Effect of particle shape on the strength and deformation mechanisms of ellipse-shaped granular assemblages. *Eng Computation*. 1995;12(2):99-108.
28. Dong Y, Cao S, Cheng X, Liu J, Cao H. Grain-size reduction of feldspar and flow of deformed granites within the Gaoligong shear zone southwestern Yunnan, China. Science China Earth Sciences, 2019. https://doi.org/10.1007/s11430-018-9351-8.
29. Pratt, W.K. Digital Image Processing. Wiley and Sons. New York, 1978.
30. Hurley R C, Herbold E B, Pagan D C. Characterization of the crystal structure, kinematics, stresses and rotations in angular granular quartz during compaction. *J Appl Cryst*. 2018; 51(4):1021-1034.
31. Haile J M, Johnston I, Mallinckrodt A J, McKay S. Molecular dynamics simulation: elementary methods. *Comput Phys*. 1993;7(6): 625-625.
32. Coutinho Y A, Rooney S C K, Payton E J. Analysis of EBSD Grain Size Measurements Using Microstructure Simulations and a Customizable Pattern Matching Library for Grain Perimeter Estimation. *Metall Mater Trans A*. 2017;48(5):2375-2395.
33. Kalo K, Grgic D, Auvray C, Giraud A, Drach B, Sevostianov I. Effective elastic moduli of a heterogeneous oolitic rock containing 3-D irregularly shaped pores. *Int J Rock Mech Min*. 2017;98:20-22.
34. Xiao J. Strain Geometric Field Theory. Bei Jing, BJ: China Science Publishing & Media Ltd; 2017.
35. Gao Y, Yan W, Gao F, Wang Z, Zhang Z. Study on mechanical properties and finite deformation constitutive model of red sandstone subjected to temperature-water-mechanics coupling. *Chin J Rock Mech Eng*. 2019;38(s1):2734-47.